\documentstyle[12pt]{article}
\begin{document}
\def\beq{\begin{equation}}
\def\eeq{\end{equation}}
\def\bea{\begin{eqnarray}}
\def\eea{\end{eqnarray}}
\def\ve{\vert}
\def\vel{\left|}
\def\ver{\right|}
\def\nnb{\nonumber}
\def\ga{\left(}
\def\dr{\right)}
\def\aga{\left\{}
\def\adr{\right\}}
\def\rar{\rightarrow}
\def\nnb{\nonumber}
\def\la{\langle}
\def\ra{\rangle}
\def\ba{\begin{array}}
\def\ea{\end{array}}
\def\tep{$B \rar K \ell^+ \ell^-$}
\def\tepm{$B \rar K \mu^+ \mu^-$}
\def\tept{$B \rar K \tau^+ \tau^-$}
\def\ds{\displaystyle}
\title{{\small {\bf The Radiative Leptonic 
$B_{c}\rightarrow \tau\bar{\nu_{\tau}}\gamma$ Decay in Two Higgs 
Doublet Model }}}
\author{\vspace{1cm}\\
{\small T. BARAKAT} \thanks
{electronic address: zayd@cc.ciu.edu.tr}\\ 
{\small Near East University},\\
{\small Lefko\c{s}a, P.O. BOX 1033  Mersin 10 - Turkey } }
\date{}
\begin{titlepage}
\maketitle
\thispagestyle{empty}
\begin{abstract}
\baselineskip .8 cm
The radiative leptonic $B_{c}\rightarrow\tau\bar{\nu_{\tau}}\gamma$ decay is 
analysed in context of 2HDM. It is shown that with large values of $tan\beta$, 
the contributions of Model II to the decay rate exceeds considerably the 
Standard Model ones, while the contributions of Model I overlap with the 
Standard Model predictions.
\end{abstract}
\vspace{1cm}
\end{titlepage}
\section{Introduction}
\baselineskip .8cm
\hspace{0.6cm}The observation of the $B_{c}$-meson by the CDF collaboration 
in the channel $B_{c}\rightarrow J/\psi\ell\nu$, with ground state mass 
$B_{c}=6.4\pm 0.39\pm 0.13$ GeV, and lifetime 
$\tau (B_{c})=0.46^{+0.18}_{-0.16}$$\pm 0.03$ ps [1], has stimulated up the 
investigation of the properties of $B_{c}$-mesons theoretically, and 
experimentally on a new footing. The particular interest of this observation 
is related to the fact that, the meson ground-state with $\bar{b}c(b\bar{c})$ 
can decay only weakly, thus providing the rather unique opportunity of 
investigating weak decays in a heavy quarkonium system. Moreover, studying 
this meson could offer an unique probe to check the perturbative QCD 
predictions more precisely, and one can get essential new information about 
the confinement scale inside hadrons. The theoretical study of the 
pure-leptonic decays of $B_{c}$-meson, such as 
$B_{c}\rightarrow \ell\bar{\nu_{\ell}}$ ($\ell=e,\mu,\tau$) can be used to 
determine the leptonic decay constant $f_{B_{c}}$ [2], as well as the 
fundamental parameters in the Standard Model (SM), such as the Cabibbo- 
Kobayashi-Maskawa (CKM) matrix elements which are poorly known at 
present. Nevertheless, the well-known ''helicity suppression'' effect make an 
experimental difficulty in the measurement of purely leptonic decays of 
$B_{c}$. Although, the $B_{c}\rightarrow \tau\bar{\nu_{\tau}}$ channel is 
free of the helicity suppression, the observation of this decay is 
experimentally difficult due to the efficiency  problem for detecting the 
$ \tau$ lepton.

Recently, the radiative leptonic 
$B_{c}\rightarrow \ell\bar{\nu_{\ell}}\gamma$ decays ($\ell=e,\mu,\tau$), 
received considerable attention as a testing ground of SM and ''new physics'', 
where no helicity suppression exists any more [3-5].
  
 Among various radiative leptonic decays, the 
$B_{c}\rightarrow \tau\bar{\nu_{\tau}}\gamma$ decay provokes special 
interest, since the SM predictions has been exploited to establish a bound on 
the branching ratio of the above mentioned decay of order $\approx 10^{-5}$ 
[5,6], and therefore can be potentially measurable in the up coming LHC 
B-factories, where the number of $B_{c}$-mesons that will be produced are 
estimated to be $\approx 2.0\times 10^{12}$ [7,8]. This will provide an 
alternative way to determine the decay constant $f_{B_{c}}$ and the CKM 
matrix elements. The decay $B_{c}\rightarrow \tau\bar{\nu_{\tau}}\gamma$ 
receive two types of contributions: 
internal bremsstrahlung (IB), and structure-dependent (SD) parts. The IB 
contributions are still helicity suppressed, while the SD ones contain the 
electromagnetic coupling constant $\alpha$ but they are free of helicity 
suppression. Therefore, the radiative decay rates of 
$B_{c}\rightarrow \ell\bar{\nu_{\ell}}\gamma$ could have an enhancement with 
respect to the purely leptonic modes of 
$B_{c}\rightarrow \ell\bar{\nu_{\ell}}$ due to the SD contributions, thus it 
enable to establish "new physics" beyond the standard model.

In this work, we will study the radiative leptonic 
$B_{c}\rightarrow \tau\bar{\nu_{\tau}}\gamma$ decay in the framework of the 
two-Higgs doublet model (2HDM) [9-11] at large $tan\beta$. The so-called 
Model I and Model II are considered, which are differ only in the 
couplings of the charged Higgs bosons to fermions. Subsequently, this paper 
is organized as follows:
 in section 2, the theoretical formalism relevance for the 
$B_{c}\rightarrow \tau\bar{\nu_{\tau}}\gamma$ decay in 2HDM is presented. 
Section 3, is devoted to the numerical analysis and the discussion of the 
results. 
     
\section{Formalism for the $B_{c}\rightarrow \tau\bar{\nu_{\tau}}\gamma$ 
decay}
\hspace{0.6cm} In the Standard Model (SM), the decay 
$B_{c}\rightarrow \tau\bar{\nu_{\tau}}\gamma$ can be studied to a very 
good approximation in terms of four-fermion interactions. The effective 
Hamiltonian relevant to the process $B_{c}\rightarrow \tau\bar{\nu_{\tau}}$ 
is:
\begin{eqnarray}
H_{eff}&=&\frac{G_{F}}{\sqrt {2}}\,a_{1}\,V_{cb}\,
\bar{c}\gamma_{\mu}(1-\gamma_{5})b
\bar{\tau}\gamma_{\mu}(1-\gamma_{5})\nu_{\tau},
\end{eqnarray}
where $G_{F}$ is the Fermi constant, $V_{cb}$ is the CKM mixing element, 
$a_{1}$ is a QCD corrected factor, which is equal $a_{1}\simeq 1.13$. 
However, in the next discussion we will put $a_{1}\simeq 1$ 

The emission of a real photon in leptonic decays of heavy mesons 
$B_{c}\rightarrow \tau\bar{\nu_{\tau}}\gamma$ can proceed via the two 
mechanisms mentioned in Sec. I. For the IB amplitude, the charged 
$B_{c}$-meson emits leptons via the axial-vector current, and the photon is 
radiated from the external charged particles. On the other hand the SD amplitude is governed by the 
vector and axial vector form factors, in which the photon is emitted from 
intermediate states. Gauge invariance leaves only two form factors 
$f_{1,2}(p^{2})$ undetermined in the SD part. The possible diagrams for this 
two mechanisms are shown in Fig. 1. Following this framework, the general form 
of the gauge invariant amplitude corresponding to Fig. 1 can be written as [6]:
\begin{eqnarray}
M(B_{c}\rightarrow \tau\bar{\nu_{\tau}}\gamma)=M_{1}+M_{2},
\end{eqnarray}
where $M_{1}$ and $M_{2}$ represent the contributions of ''inner 
bremsstrahlung'' (IB), and ''structure-dependent'' (SD) parts, given by:
\bea
 M_{1} &=& ie\frac{G_{F}}{\sqrt{2}} \,V_{cb} f_{B_c} m_\tau 
\varepsilon_{\alpha}\,\bar
u(p_1)\Bigg\{
\frac{\gamma_{\alpha} \not\! q+2p_{1\alpha}}{2 p_1. q} -\frac{p_\alpha}
{p.q} \Bigg\}\ga 1 - \gamma_5 \dr v(p_2) ~,
\eea

\bea
 M_{2} &=&ie\frac{G_{F}}{\sqrt{2}} \,V_{cb} \varepsilon_{\alpha}\Bigg\{
\frac{-i\,f_1(p^2)}{m_{B_c}^2} \epsilon_{\alpha \beta \rho \gamma}
 p_\rho q_\gamma  \nnb \\
&+& \frac{f_2(p^2)}{m_{B_c}^2}
\left[ p_\alpha q_\beta - g_{\alpha \beta} p.q \right]
\Bigg\}
\bar u(p_1) \gamma_\beta \ga 1 -\gamma_5 \dr v(p_2) ~, 
\eea
where $\varepsilon_{\alpha}$ is the photon polarization vector, $p_{1}$, 
$p_{2}$, and q are the four momenta of $\tau$, $\nu_{\tau}$, and $\gamma$, 
respectively.  $f_{B_{c}}$ is the $B_{c}$- meson leptonic decay constant, 
$f_{1,2}(p^{2})$ corresponding to parity conserving and a parity violating 
formfactors, $p=P_{B_{c}}=p_{1}+p_{2}$ is the momentum transfer to lepton 
pair. The necessary matrix elements related to the $f_{B_{c}}$, and to the 
hadron transition form factors $f_{1,2}(p^{2})$ are defined as follows [4]:
\begin{eqnarray}
\left<0\mid \bar c \gamma_{\mu} \gamma_{5} b  \mid B_{c} \right> = - i
f_{B_{c}} P_{B\mu}~,
\end{eqnarray}
\begin{eqnarray}
\left<\gamma(q)\mid \bar c \gamma_{\alpha} b  \mid B_c(p+q) \right> = e\frac{
f_{1}(p^{2})}{m^{2}_{B_{c}}}\epsilon_{\alpha \beta \rho \gamma} 
\varepsilon^\beta p^\rho q^\gamma~,
\end{eqnarray}
\begin{eqnarray}
\left<\gamma(q)\mid \bar c \gamma_{\alpha} \gamma_{5} b \mid B_c(p+q) \right> =
 -ie\frac{ f_{2}(p^{2})}{m^{2}_{B_{c}}}\varepsilon_\alpha \left
[g_{\alpha\beta} (p.q)-p_\alpha q_\beta \right].
\end{eqnarray}
We want now to consider the $B_{c}\rightarrow \tau \bar \nu_\tau \gamma$ 
decay in the context of a 2HDM with no flavor changing neutral currents 
(FCNC) allowed at the tree level, i.e. Model I and Model II. The interaction 
lagrangian of fermions with the charged Higgs fields in both models is given 
by [12]:
\bea
 L&=&\frac{g_{W}}{2\sqrt{2}M_{W}} \Bigg \{ V_{ij}m_{u_{i}}X
\bar{u_{i}}(1-\gamma_{5})d_{j}+V_{ij}m_{d_{j}}Y
\bar{u_{i}}(1+\gamma_{5})d_{j} \nnb \\
&+&m_{\ell}Z\bar{\nu}(1+\gamma_{5})\ell \Bigg \} H^{\pm}+h.c.,
\eea
where $g_{W}$ is the weak coupling constant, $M_{W}$ is the $W$- boson mass, 
$H^{\pm}$ is the charged physical field, and $V_{ij}$ is the relevant 
elements of CKM matrix. In model I, $X\,=\,cot\beta$, $Y\,=Z\,=\,-cot\beta$ 
and in model II, $X\,=\,cot\beta$, and $Y\,=\,Z=\,\,tan\beta$.

The decay $B_c \rightarrow \tau \bar \nu_\tau \gamma$ in 2HDM proceeds through 
the same Feynman diagrams (which are displayed in Fig. 1) that mediate the 
process in SM, except the $W$-boson is replaced by the charged scalar Higgs 
boson $H^{\pm}$, i.e. $(W \rightarrow H^{\pm})$.
\vspace{11.0cm} 
\begin{center}
Fig. 1
\end{center}

The matrix element corresponding to the diagram $(W \rightarrow H^{\pm})$ 
where the photon is radiated from $\tau$ lepton, (see fig.1) is:
\begin{eqnarray}
M_{1}^{2HDM}&=&i\, e \,\frac{G_{F}}{\sqrt{2}} \,V_{cb} f_{B_{c}}
\varepsilon_\alpha\frac{m_{\tau}m^{2}_{B_{c}}}{M^{2}_{H}(m_{b}+m_{c})}
(m_{b}ZY-m_{c}ZX) \times \nonumber \\
&&\bar u(p_{1}) \left\{\frac{2 p_{1\alpha} + \gamma_\alpha \not\!q}{2 p_{1} q}
 \right\}(1 - \gamma_{5} ) v(p_{2})~,
\end{eqnarray}
where $f_{B_{c}}$ is the leptonic decay constant of $B_{c}$ meson, defined as:
\begin{eqnarray}
\left<0\mid \bar c  \gamma_{5} b  \mid B_{c}(p+q) \right> = 
-if_{B_{c}} \frac{m^{2}_{B_{c}}}{(m_{b}+m_{c})}.
\end{eqnarray} 

While the contribution of the structure dependent part to the $B_{c}
\rightarrow \tau \bar \nu_{\tau} \gamma$ decay, i.e., when photon is radiated
from initial quark lines, due to the charged Higgs exchange can be obtained 
by considering the following correlation function:
\begin{eqnarray}
M^{SD}_{\alpha}&=& -ie\frac{G_{F}}{\sqrt{2}} V_{cb}\varepsilon_{\alpha}
Z\frac{m_{\tau}}{M^{2}_{H}}   \int d^{4} x e^{i q x}\times \nonumber \\
&&\!\! <0 \mid T \Bigg \{ \Bigg [\bar c(0)\Big (m_{b}Y (1+ \gamma_{5}) 
+m_{c}X (1 - \gamma_{5})\Big ) b(0)\Bigg ] J_{\alpha}^{el} (x) \Bigg \}
\mid B_{c}> \times \nonumber \\
&&\bar u(p_1) \gamma_\beta \ga 1 -\gamma_5 \dr v(p_2),
\end{eqnarray}
where $J_{\alpha}^{el}(x)$ is the 
electromagnetic current for $b$ or $c$ quarks. The hadronic matrix elements 
involving the scalar and pseudoscalar currents in Eq.(11) are parameterized 
such that:
\begin{eqnarray}
\int d^{4} x e^{i q x} 
<0 \mid T ( \bar c(0) \gamma_{5} b(0) J_{\alpha}^{el} (x)
 )\mid B_{c}(p+q)>=f_{B_{c}}\frac{m^{2}_{B_{c}}}{(m_{b}+m_{c})}
\frac{p_{\alpha}}{p.q},
\end{eqnarray}
\begin{eqnarray}
\int d^{4} x e^{i q x} 
<0 \mid T ( \bar c(0)\, b(0) J_{\alpha}^{el} (x))\mid B_{c}(p+q)>=0.
\end{eqnarray}
The parameterization of the hadronic matrix elements given in Eqs. (12,13) are 
particularly well suited for our purposes since in the 2HDM, the vertex 
$b(c)\sim (1-\gamma_{5})$ or $b(c)\sim (1+\gamma_{5})$, hence, the vector 
part of this correlator is zero, and the active part of this correlator is 
the axial-part given by: 
\begin{eqnarray}
 M^{A(SD)}_{\alpha}&=& -ie\frac{G_{F}}{\sqrt{2}} \,V_{cb}f_{B_{c}}
\varepsilon_\alpha\frac{m_{\tau}m^{2}_{B_{c}}}{M^{2}_{H}}
\frac{(m_{b}ZY -m_{c}ZX)}{(m_{b} +m_{c})}\times  \nonumber \\
&&\bar u(p_{1})
\frac{ p_{\alpha}}{ p q}(1 - \gamma_{5}) v(p_2)~,
\end{eqnarray}
and the total matrix element in 2HDM becomes:
\bea
 M_{2}^{2HDM}(B_{c}\rightarrow \tau \bar \nu_{\tau} \gamma) &=& 
ie\frac{G_{F}}{\sqrt{2}} \,V_{cb} f_{B_c} \varepsilon_{\alpha}\,
\frac{ m_\tau m^{2}_{B_{c}}}{M^{2}_{H}}
\frac{(m_{b}ZY -m_{c}ZX)}{(m_{b} +m_{c})} \times \nonumber \\
&&\bar u(p_1)\Bigg\{
\frac{\gamma_{\alpha} \not\! q+2p_{1\alpha}}{2 p_1. q} -\frac{p_\alpha}
{p.q} \Bigg\}\ga 1 - \gamma_5 \dr v(p_2) ~.
\eea

At this accuracy it is easy to check that the modified total amplitude for the 
radiative leptonic B-decays of $B_c \rightarrow \tau \bar \nu_\tau \gamma$ 
is gauge invariant:
\begin{eqnarray}
M_{(total)}(B_{c}\rightarrow \tau\bar{\nu_{\tau}}\gamma)=M^{new}_{1}+M_{2},
\end{eqnarray}
where
\bea
 M^{new}_{1} &=& ie\frac{G_{F}}{\sqrt{2}} \,V_{cb} f_{B_c} m_\tau 
\varepsilon_{\alpha}\ C^{2HDM}\bar u(p_1)\Bigg\{
\frac{\gamma_{\alpha} \not\! q+2p_{1\alpha}}{2 p_1. q} -\frac{p_\alpha}
{p.q} \Bigg\}\ga 1 - \gamma_5 \dr v(p_2) ~.
\eea
Therefore, in this model the charged Higgs contribution 
modifies only the so-called $M_{1}$ part of the SM, and it does not induce 
any new contribution to the so-called $M_{2}$ (see Eq.3): 
\bea
C^{2HDM}=\Bigg\{ \frac{m^{2}_{B_{c}}}{M^{2}_{H}}
\frac{(m_{b}ZY -m_{c}ZX)}{(m_{b} +m_{c})}+1 \Bigg \}.
\eea

The 2HDM is sensitive to two basic free parameters, namely $tan\beta$, and 
the charged Higgs mass $M_{H}$. If we formally set $Z \rightarrow 0$ in 
Eq. (18), the resulting expression is expected to 
coincide with the $B_c \rightarrow \tau \bar\nu_{\tau}\gamma$ decay, which 
was investigated in the framework of SM [6].

After lengthy, but straightforward calculation for the squared
matrix element, we get:
\begin{eqnarray}
{\mid  M_{total} \mid}^2 = {\mid  M^{new}_1 \mid}^2 +
2 \,\mbox{\rm Re}\left[  M^{new}_1  M_2^\dagger \right]
+ {\mid  M_2 \mid}^2~,
\end{eqnarray}
where
\begin{eqnarray}
{\mid  M^{new}_1 \mid}^2 &=& \frac{G^2}{2}{\mid V_{cb} \mid}^2 e^2
( - 4 f_{B_c}^2 m_\tau^2){\mid C^{2HDM} \mid}^{2}
\frac{1}{(p_1 q)^2 (p q)^2} \nonumber \\
&\times& \!\!\Bigg\{ 2 p^2 (p_1 p_2) (p_1 q)^2 + (p q)^2 \left[
(p_1 p_2) (2 m_\tau^2 - p_1 q) +
(p_2 q) ( m_\tau^2 - 2 p_1 q) \right] \nonumber \\
&+& \!\! (p q) (p_1 q) \left[ (p p_2) (p_1 q) - (p p_1)
(4 p_1 p_2 + p_2 q) \right] \Bigg\}~,
\\ \nonumber \\ \nonumber \\
2 \,\mbox{\rm Re}\left[  M^{new}_1  M_2^\dagger \right] &=&
\frac{G^2}{2} {\mid V_{cb} \mid} ^2 e^2 (- 16 f_{B_c} m_\tau^2)\,C^{2HDM}
\frac{1}{(p_1 q) (p q)}\nonumber \\
&\times& \!\! \Bigg\{ \frac{ f_2(p^2)}{m_{B_c}^2} \, p^2 (p_1 q)
(p_2 q)
+ (p q)^2 \left[ \frac{ f_2(p^2)}{m_{B_c}^2} \,(p_1 p_2 + p_2 q) -
\frac{ f_1(p^2)}{m_{B_c}^2} \, (p_2 q) \right] \nonumber \\
&-& \!\! \frac{ f_2(p^2)}{m_{B_c}^2}
\left[ (p p_2) (p_1 q) + (p p_1) (p_2 q) \right] \Bigg\}~,
\\ \nonumber \\ \nonumber \\
{\mid  M_2 \mid}^2 &=& \frac{G^2}{2} {\mid V_{cb} \mid}^2 e^2
16\left[ \frac{{\mid f_1(p^2) \mid}^2}{m_{B_c}^4} +
\frac{{\mid f_2(p^2) \mid}^2}{m_{B_c}^4} \right] \nonumber \\
&\times& \!\! \left\{ (p p_2) (p q)  (p_1 q) +
(p_2 q) \left[ (p p_1)  (p q) - p^2 (p_1 q) \right]
\right\}~.
\end{eqnarray}
All calculations have been performed in the rest frame of the
$B_c$ meson.
The dot products of the four--vectors are defined if the photon
and neutrino
(or electron) energies are specified. The Dalitz boundary for the
photon energy
$E_\gamma$ and neutrino energy $E_2$ are defined as follows:
\begin{eqnarray}
\frac{m_{B_c}^2 - 2 m_{B_c} E_\gamma - m_\tau^2}
{2 m_{B_c}} \le \!\!\! &E_2& \!\!\! \le
\frac{m_{B_c}^2 - 2 m_{B_c} E_\gamma - m_\tau^2}
{2 (m_{B_c} - 2 E_\gamma)}~, \nonumber \\ \nonumber \\
0 \le \!\!\! &E_\gamma& \!\!\! \le
\frac{m_{B_c}^2 - m_\tau^2}{2 m_{B_c}}~.
\end{eqnarray}
 It is now straightforward to work out the expression for the differential 
decay rate in the lepton and photon energies:
\begin{eqnarray}
\frac{d\Gamma}{d E_2\,d E_\gamma} = \frac{1}{64 \pi^3 m_{B_c}}
{\mid  M_{total} \mid}^2~.
\end{eqnarray}

The differential $( d\Gamma/d E_\gamma)$ and total decay
width
are singular at the lower limit of the photon energy, and this
singularity
which is present only in the ${\mid  M^{new}_1 \mid}^2$ contribution
is due to the soft photon emission from charged lepton line.
On the other hand, ${\mid  M_2 \mid}^2$ and
$\mbox{\rm Re}\left[  M^{new}_1  M_2^\dagger \right]$ terms
are free from this singularity.
In this limit the $B_c \rightarrow \tau \bar \nu_\tau \gamma$ decay can
not be distinguished from the $B_c \rightarrow \tau \bar \nu_\tau$ decay. 
Therefore, in order to
obtain a finite result for the decay width, we must consider both
decays together. The infrared singularity arising in the 
${\mid M^{new}_1  \mid}^2$ contribution must be canceled with $O(\alpha)$ 
virtual correction to the $B_c \rightarrow \tau \bar \nu_\tau$ decay. In the 
SM this cancellation explicitly was shown in [13].
 
In this work, the $B_c \rightarrow \tau \bar \nu_\tau \gamma$ process is not 
considered as a $O(\alpha)$ correction to the 
$B_c \rightarrow \tau \bar \nu_\tau$ decay, but rather a separate decay 
channel with hard photon radiation. Therefore we
impose a cut off value on the photon energy, which will set an
experimental limit on the minimum detectable photon energy. We consider 
the case for which the photon energy threshold is larger than
$50$, MeV
i.e., $E_\gamma \ge a \,m_{B_c}$, where $a \ge 0.01$. An integration over all 
the possible values of the lepton energy $E_{2}$ gives the total decay width 
as a function of the photon energy:
\\
\\

\begin{eqnarray}
\Gamma &=& \frac{G^2 \alpha m_{B_c}^3}{64 \pi^2} {\mid V_{cb}
\mid}^2
\Bigg\{ 4 f_{B_c}^2{\mid C^{2HDM} \mid}^{2} \int_\delta^{1-r} dx 
\,\frac{r}{x(1-x)}
\Bigg[ -4 + 8 r - 4 r^2 + 10 x - 14 r x \nonumber \\
&+& 4 r^2 x - 9 x^2 + 7 rx^2+ 3 x^3 
+(1-x) (2 -2 r^2 - 3 x + rx + 2 x^2 )
\ell n (\frac{1-x}{r}) \Bigg]\nonumber \\
&-& \!\! 4 f_{B_c} \, C^{2HDM}\int_\delta^{1-r} dx \,\frac{r x}{1-x} \Bigg[
(1-r-x) \Big( f_1(x) x + f_2(x) (1+r-2 x) \Big) \nonumber \\
&-& \!\! (1-x) \Big( f_1(x) x + f_2(x) (2r-x) \Big)
\ell n (\frac{1-x}{r}) \Bigg]\nonumber \\
&+&\!\!\frac{1}{3} \int_\delta^{1-r} dx
\left[ {\mid f_1(x) \mid}^2 + {\mid f_2(x) \mid}^2 \right]
\frac{1}{(1-x)^2} \, x^3 (2 + r - 2 x) (1-r-x)^2 
\Bigg\},
\end{eqnarray}
where $x = 2 E_\gamma/m_{B_c}$is the dimensionless photon energy,
$r = m_\tau^2/m_{B_c}^2$ and $\delta = 2 a$.

\section{NUMERICAL ANALYSIS}
 To calculate the decay width, explicit forms of the form factors 
$f_1$ and $f_2$ are needed. These form factors were calculated in the 
framework of the light-front quark model in [14], and in the light-cone QCD 
sum rules [3], where in [3] it was found that, the best agreement 
is achieved by the following pole forms for the form factors:
\bea
f_1(p^2) = \frac{f_1(0)}{1 - p^2/m_1^2}~,~~~~~~~~~~
f_2(p^2) = \frac{f_2(0)}{1 - p^2/m_2^2}~,
\eea
where
\bea
f_1(0) = 0.44 \pm 0.04~ \mbox{\rm GeV}~, ~~~~~ m_1^2 = 43.1
~\mbox{\rm GeV}^2~,\nnb \\
f_2(0) = 0.21 \pm 0.02~ \mbox{\rm GeV}~, ~~~~~ m_2^2 = 48.0
~\mbox{\rm GeV}^2~.\nnb
\eea
 On the other hand, in evaluating the decay width, we have used the following 
set of parameters: $G_{F}=1.17{~}.10^{-5}~GeV^{-2}$, $\alpha =1/137$, 
$m_{b}=4.8\,GeV$, $m_{c}=1.4\,GeV$, $m_{B_{c}}=6.3\,GeV$, 
$m_{\tau}=1.78\,GeV$, $f_{B_c}=0.36$\,GeV [15], 
$V_{cb}=0.04$ [16] and,  $\tau (B_c) = 0.46 \times 10^{-12}~s$ [1].

In this regard we should also recall that the free parameters of the 2HDM 
model namely $tan\beta$, and $M_{H}$ are not arbitrary, but 
there are some semiquantitative restrictions on them using the existing 
experimental data. The most direct bound on the charged Higgs boson mass 
comes from top quark decays, which yield the bound $M_{H}\,>\,147 GeV$ for 
large $tan\beta$ [17]. For pure type-II 2HDM's one finds $M_{H}\,>\,300 GeV$, 
coming from the virtual Higgs boson contributions to 
$b\rightarrow s\gamma$ [18]. Furthermore, there are no experimental upper 
bounds on the mass of the charged Higgs boson, but one generally expects to 
have $M_{H}< 1 TeV$ in order that perturbation theory remain valid [19]. For 
large $tan\beta$ the most stringent constraints on $tan\beta$ and $M_{H}$ are 
actually on their ratio, $tan\beta /M_{H}$. The current limits come from the 
measured branching ratio for the inclusive decay 
$B\rightarrow X\tau \bar \nu$, giving $tan\beta /M_{H} < 0.46\,GeV^{-1}$ [20], 
and from the upper limit on the branching ratio for 
$B\rightarrow \tau \bar \nu$, giving $tan\beta /M_{H} < 0.38\,GeV^{-1}$ [21].

For illustrative purposes we consider four values of $tan\beta$, namely 
$tan\beta=$5, 10, 30 and 50 and let $m_{H^{\pm}}=150\,GeV$. Then we consider 
two values of $m_{H^{\pm}}$, namely $M_{H^{\pm}}$=200, 400 GeV, and we allow 
$tan\beta$ to range between 0 to 60. The results of this numerical analysis 
are graphically shown in figures 2-4. In these figures the differences 
between the 2HDM's and the SM are shown for two different fixed cut off   
 values, i.e., $\delta = 0.016$ and $\delta = 0.032$ both for Model I and 
Model II. 

The results for the differential decay branching ratio 
dBR($B_{c}\rightarrow \tau \bar \nu \gamma)/dx$ as a function of 
$x=2E_{\gamma}/m_{B_{c}}$ 
for different values of $tan\beta$, $M_{H}=150\,GeV$ are presented in 
Fig.2, while the branching ratio (BR) for $B\rightarrow \tau \bar \nu \gamma$ 
decay is shown in figure 3-4 as a function of $M_{H}$ for various values of 
$tan\beta$, and as a function of $tan\beta$ for different values of 
$M_{H}$. Results are shown for Model I, and Mode lI.  

It is observed that Model II gives both a bigger differential decay branching 
ratio, and a bigger  branching ratio than the SM rates of (up to 
three orders of magnitude [$M_{H}=150\,GeV$]) for large values 
of $tan\beta>\,20$, while for small values of $tan\beta<\,10$ results 
approaches its SM value. \\
In model I the situation is somewhat totally different. Curves overlap 
with the SM results all the way. This behavior obviously reflects the 
$H^\pm$ fermion couplings, which are proportional to $cot\beta$ in this model.
  
In conclusion, this study shows that the branching ratios for 
$B_{c}\rightarrow \tau \bar \nu \gamma$ could be at the level of $10^{-4}$ in 
the 2HDM, which may be detectable at the ongoing LHC. When enough $B_{c}$ 
events are collected, this decay will be able to provide alternative channel 
to extract new restrictions for the free parameters $tan\beta$ and 
$M_{H^{+}}$ of the 2HDM model.

\clearpage
\begin{center}
Figure Captions
\end{center}
~~~~\\
Figure 1. : The relevant Feynman diagrams, responsible for
             $B_{c}\rightarrow \tau \bar \nu \gamma$ decay. \\
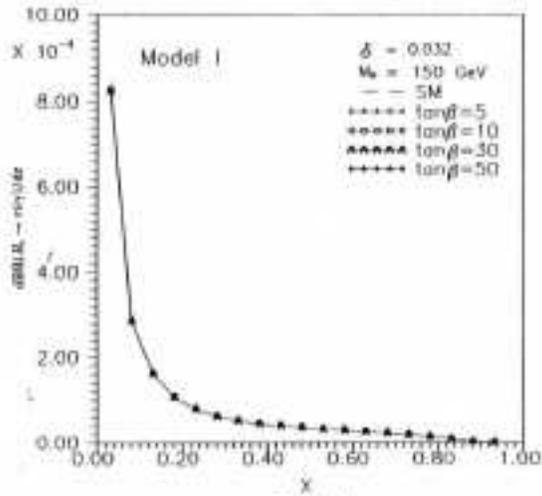
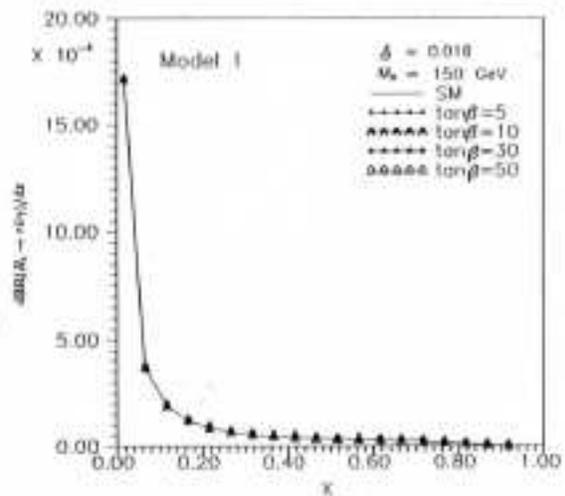
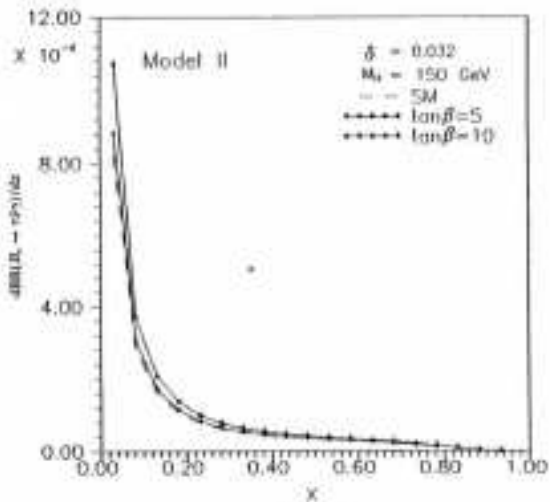
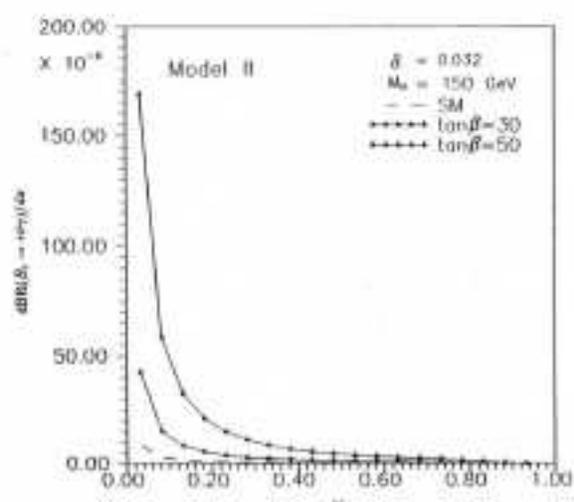
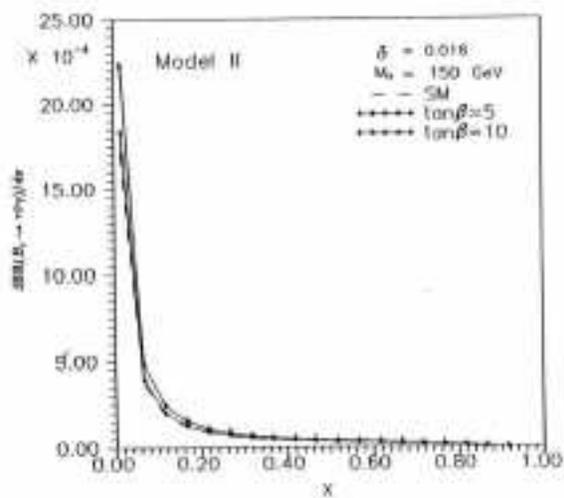
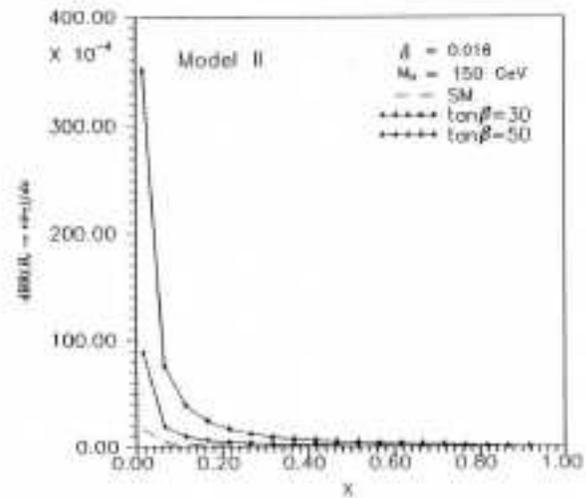
Figure 2. : The dependence of the differential Branching ratio of 
             dBR($B_{c}\rightarrow \tau \bar \nu \gamma)/dx$ as a function of 
             the photon energy $x=2E_{\gamma}/m_{B_{c}}$ for both models 
              Model I, and Model II.  \\
Figure 3. : The dependence of the Branching ratio on the charged Higgs 
             boson mass at different values of $tan\beta$ for both models 
              Model I, and Model II.\\
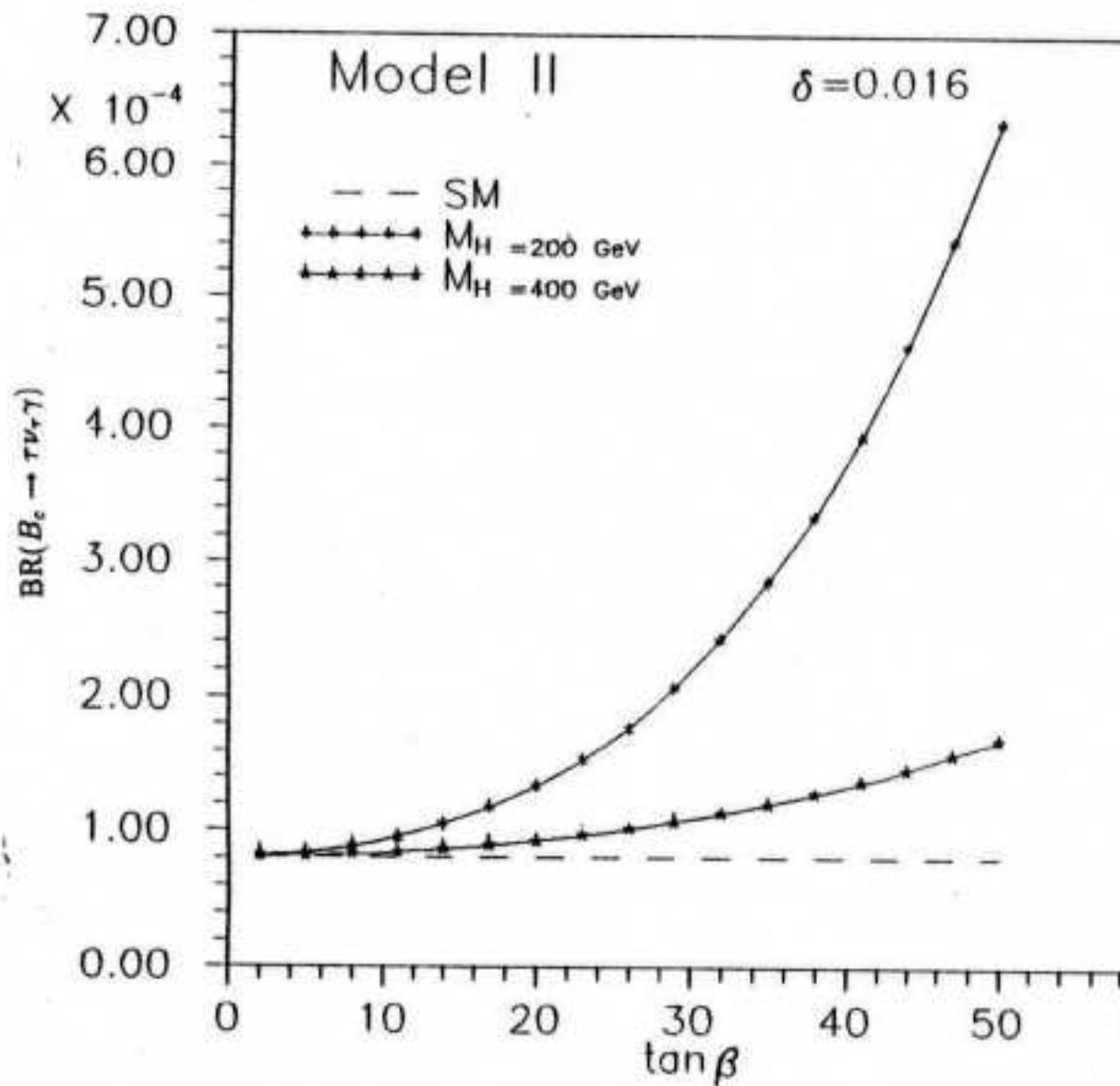
Figure 4. :  The dependence of the Branching ratio on $tan\beta$ 
             at different values of the charged Higgs boson mass for both 
             models Model I, and Model II.\\    
\pagebreak

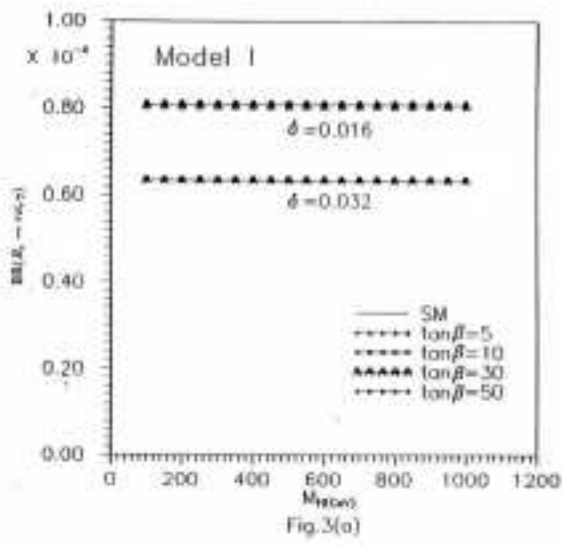
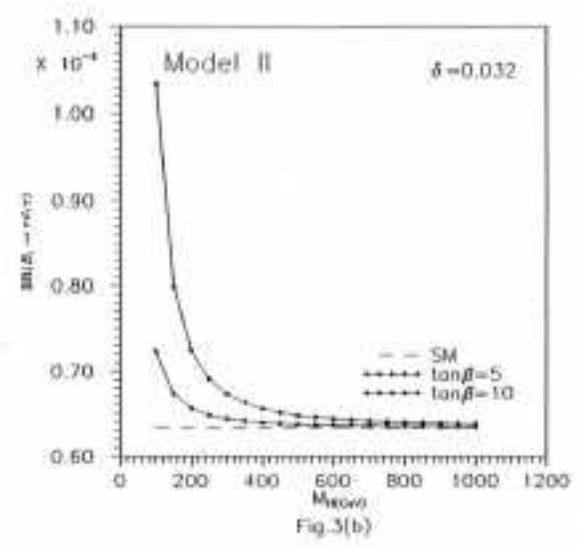
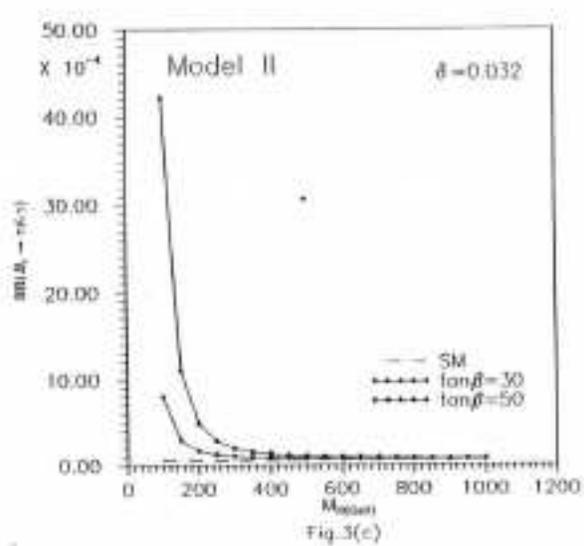
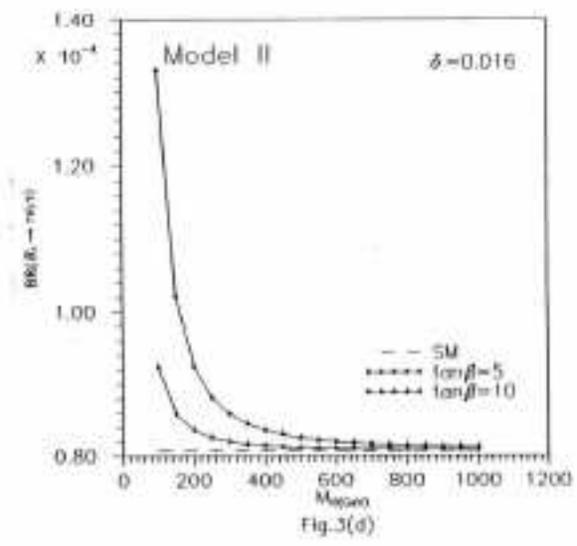
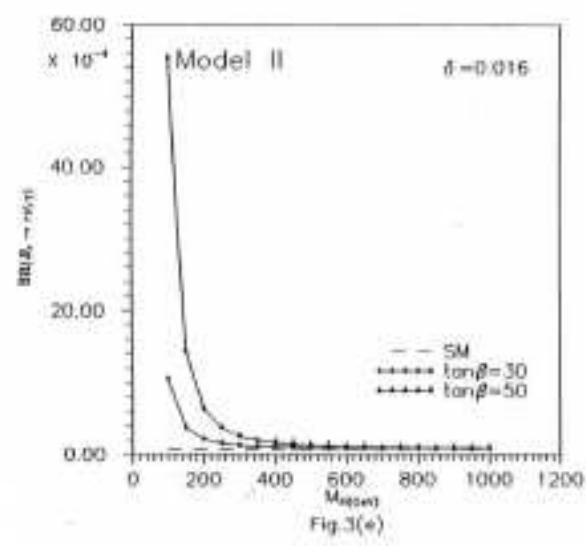
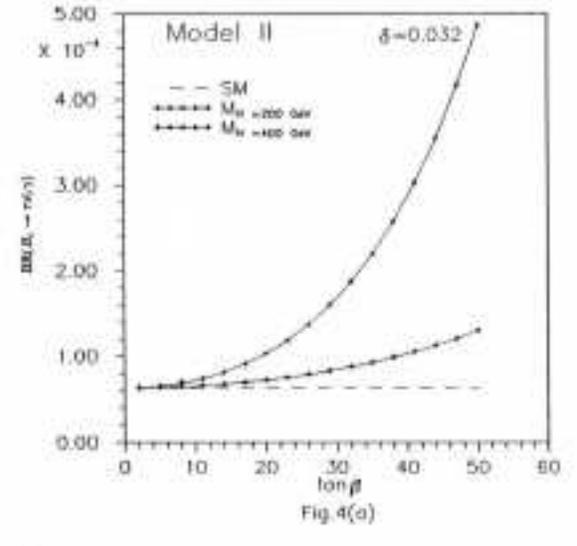
\end{document}